\documentstyle[12pt,graphicx,amsmath]{article}
\begin{document}
\title
{
The Impact Crater Size-Frequency Distribution on 
Pluto Follows a Truncated Pareto Distribution: 
Results from a First Data Set 
Based on the Recent New Horizons' Flyby
}
\author
{L. Zaninetti             \\
Dipartimento di Fisica , \\
           Via Pietro Giuria 1,   \\
           10125 Torino, Italy   \\ \\
F. Scholkmann                    \\
Research Office of Complex Physical and Biological 
Systems (ROCoS),   \\
Bellariarain 10, \\
8038 Z\"{u}rich, Switzerland 
}
\maketitle
\section*{}
Recently it could be shown (Scholkmann, \textit{Prog. in Phys.}, 
2016, v.\,12(1), 26-29) that the impact crater 
size-frequency distribution of Pluto 
(based on an analysis of first images obtained by the recent 
New Horizons’ flyby) follows a power 
law ($\alpha = 2.4926 \pm 0.3309$) in the interval 
of diameter ($D$) values ranging from $3.75 \pm 1.14$~km 
to the largest determined value of $37.77$~km. 
A reanalysis of this data set revealed that the whole crater 
SFD (i.e., with values in the interval of 1.2--37.7~km) 
can be described by a truncated Pareto distribution.

\section{Introduction}
The recent flyby from NASA's New Horizons spacecraft has allowed the obtain high-resolution images of Pluto's surface morphology and thus enabled a first determination of the impact crater size-frequency distribution (SFD) of a specific region, i.e., covering parts of Pluto’s regions \textit{Sputnik Planum}, \textit{Al-Idrisi Montes} and \textit{Voyager Terra} \cite{Scholkmann2016}.

The first analysis of the crater SFD used a power law of the type $p(x) \sim x^{-\alpha}$ to model the data. In the present paper we show the results of an extended analysis.
The inverse power law scaling is known as the Pareto distribution which scales as  $p(x) \sim x^{-(c+1)}$. The hypothesis we tested in the present paper was if a upper truncated Pareto distribution (i.e., a Pareto distribution in which the probability range is limited rather than infinite) can improve the modelling of the empirical crater SFD presented in \cite{Scholkmann2016}.

We review the properties of the Pareto and the truncated Pareto distributions in Section \ref{secpareto}, and report in Section \ref{secapplications} the results of applying the truncated Pareto distribution to the novel Pluto crater SFD 
data set.

\section{From the Pareto to the truncated Pareto distribution}
\label{secpareto}
In the follwing we report the definitions of the probability density function (PDF), the distribution function (DF), the survival function (S) and the maximum likelihood estimator (MLE)
for the two distributions here analyzed. The sample is made by $n$ crater diameter $(D)$ values denoted by $x_i$.

\subsection{The Pareto distribution}

The  Pareto PDF is given by
\begin {equation}
f(x;a,c) = {c a^c}{x^{-(c+1)}},
\label{pareto}
\end {equation}
with $ c~>0$; the Pareto DF is defined as
\begin{equation}
F (x;a,c)=1-a^cx^{-c},
\end{equation}
and the survival function is given by
\begin{equation}
S(x;a,c) = 1 -F (x;a,c).
\end{equation}
The parameter values can be estimated by applying the MLE:
\begin{subequations}
\begin{align}
a          & = \min (x_i),  \\
\frac{1}{c}& = \left( \frac{1}{n} \right)
\displaystyle\sum_{i=1}^n \ln \left(\frac{x_i}{\tilde a} \right) \, .
\end{align}
\end{subequations}
More details can be found in \cite{evans}.

\subsection{The truncated Pareto distribution}
An upper truncated Pareto random variable is defined in the interval
$[a, b]$, and the PDF is given by
\begin {equation}
\label{pdfparetotruncated}
f_T(x;a,b,c ) = \frac{ca^cx^{-(c+1)}}{1-\left (\frac{a}{b}\right)^c} \,;
\end {equation}
and the truncated DF is defined as
\begin{equation}
F_T(x;a,b,c) =
\frac {1 -\left(\frac {a}{x}\right)^c} {1-\left(\frac{a}{b}\right)^c} \, .
\end{equation}
The MLE determines the parameters according to
\begin{subequations}
\begin{align}
a          & = \min (x_i),  \\
b          & = \max (x_i), \\
0          & = \frac {n}{{\tilde c}} +
\frac {n  \left(\frac {a}{b}\right)^{\tilde c}
\ln \left(\frac{a}{b}\right)}{ 1-\left(\frac{a}{b}\right)^{\tilde c }}
- \displaystyle\sum_{i=1}^n  [\ln x_i-\ln a] \, ,
\end{align}
\end{subequations}
where the value of $\tilde c$ can be found using the Brent's method
to find a root of a nonlinear function,
i.e. by applying the FORTRAN subroutine ZBRENT \cite{press}.
More details can be found in \cite{Zaninetti2008d}.

\section{Data analysis and results}

\label{secapplications}
For statistical testing the Kolmogorov--Smirnov (K-S) test \linebreak \cite{Kolmogoroff1941,Smirnov1948,Massey1951} was employed which does not  require data binning. The K-S test, as implemented by the FORTRAN subroutine KSONE \cite{press}, finds the maximum  distance $(d_{max})$ between the theoretical and the empirical DF as well the  significance  level  $P_{KS}$ (see equations  14.3.5 and 14.3.9  in \cite{press}). A values of $ P_{KS} \geq 0.1 $ assures that the fit is acceptable.

\begin{figure}
{
\includegraphics[width=8.5cm]{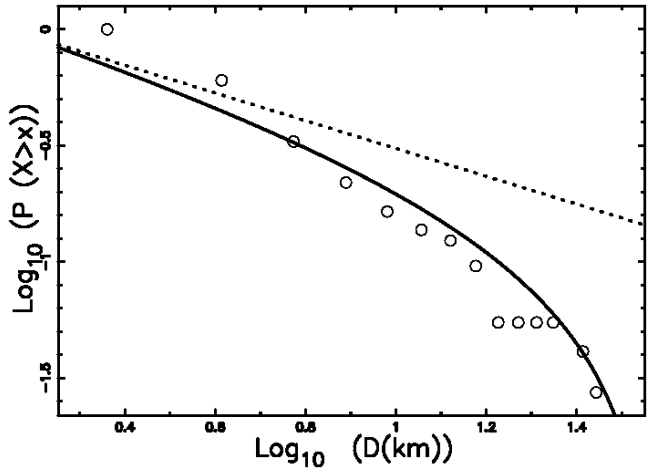}
}
\caption{
Survival function (S) in a log-log plot  for crater size in $D = [1.38$~km, 37.77~km]. Empty circles: empirical data, full line: S of the truncated Pareto  PDF, dotted line: S of the Pareto  PDF. The K-S test for the truncated Pareto gave $ P_{KS} = 0.128$ and $d_{max} = 0.134$. The K-S test for the  Pareto
gave $ P_{KS}= 0.0075$ and $d_{max}=0.192$.
}
\label{pareto_tronc_craters}
\end{figure}

When using the impact crater SFD data of Pluto with $D = [3.75$~km, 37.77~km] the Pareto PDF gave $c = 1.5299$ and thus $\alpha = 2.5299$ (similar to the value $\alpha = 2.4926$ reported by \cite{Scholkmann2016}), and the K-S test gave $ P_{KS} = 0.866$ and $d_{max} = 0.091$. Figure \ref{pareto_tronc_craters} shows the empirical and and the two fitted distributions when the interval of crater size values is enlarged so that all $D$ values are included in the fitting, i.e., $D = [1.2$~km, 37.77~km]. It can be clearly recognized that the truncated Pareto distribution describes the empirical crater SFD quite well over the whole interval of $D$ values available.

\section{Conclusions}
\markboth{Zaninetti and Scholkmann. The Impact Crater Size-Frequency Distribution on Pluto follows a truncated Pareto Distribution}{\thepage}

The distribution of crater diameters of planets is commonly modeled by a power law. A small modification of the "simple" PDF by a truncated Pareto PDF (as  given by equation (\ref{pdfparetotruncated})) allows to avoid the dichotomy of the infinite rather than finite range of existence and it provides a better K-S test \linebreak statistics with respect to the Pareto PDF (i.e., a "simple" \linebreak power law), see captions of Figure \ref{pareto_tronc_craters}.

In conclusion, we could show that the empirical impact crater SFD of Pluto (using a first data set based on recent New Horizons' flyby) is in good agreement with a truncated Pareto distribution. Applying the same modelling approach to an extended data set of Pluto's crater values is warranted to confirm our results -- a task to be done as soon as new images of the New Horizon spacecraft are available.


\end{document}